\documentclass[letterpaper, 10 pt, conference]{ieeeconf}  
\usepackage{fancyhdr}
\usepackage[normalem]{ulem}
\usepackage[hyphens]{url}
\usepackage{hyperref}
\usepackage{multirow}
\usepackage{siunitx}
\usepackage{amssymb}
\usepackage{tabularx}
\usepackage{subfig}

\usepackage{amsmath}
\usepackage{mathtools}
\usepackage{dcolumn}
\usepackage{array}
\usepackage{mdwmath}
\usepackage{mdwtab}
\usepackage{eqparbox}
\usepackage{multirow} 
\usepackage{dblfloatfix}
\usepackage{url}
\usepackage{enumerate}
\usepackage{graphicx}

\usepackage{algorithm,algcompatible,amsmath}
\algnewcommand\INPUT{\item[\textbf{Input:}]}%
\algnewcommand\OUTPUT{\item[\textbf{Output:}]}%

\IEEEoverridecommandlockouts                              
\title{\LARGE \bf
\hspace{5ex} D\textsuperscript{3}NOC: Dynamic Data-Driven Network On Chip\newline in Photonic Electronic Hybrids
}

\author{Armin Mehrabian$^{1}$, Shuai Sun$^{1}$, Vikram K. Narayana$^{1}$, Volker J. Sorger$^{1}$, and Tarek El-Ghazawi$^{1}$ 
\thanks{$^{1}$The authors are affiliated with the Electrical and Computer Engineering department of the George Washington University, Washington DC, USA\protect\\
Email: armin@gwu.edu }%
}

\begin{document}

\maketitle
\thispagestyle{empty}
\pagestyle{empty}

\begin{abstract}

In this paper we present a reconfigurable hybrid Photonic-Plasmonic Network-on-Chip (NoC) based on the Dynamic Data Driven Application System (DDDAS) paradigm. In DDDAS computations and measurements form a dynamic closed feedback loop in which they tune one another in response to changes in the environment. Our proposed system enables dynamic augmentation of a base electrical mesh topology with an optical express bus during the run-time. In addition, the measurement process itself adjusts to the environment. In order to achieve lower latencies, lower dynamic power, and higher throughput, we take advantage of a Configurable Hybrid Photonic Plasmonic Interconnect (CHyPPI) for our reconfigurable connections. We evaluate the performance and  power of our system against kernels from NAS Parallel Benchmark (NPB) in addition to some synthetically generated traffic. In comparison to a 16$\times$16 base electrical mesh, D\textsuperscript{3}NOC shows  up to 89\% latency and 67\% dynamic power net improvements beyond overhead-corrected performance. It should be noted that the design-space of NoC reconfiguration is vast and the goal of this study is not design-space exploration. Our goal is to show the potentials of adaptive dynamic measurements when coupled with other reconfiguration techniques in the NoC context.

\end{abstract}

\section{INTRODUCTION}

As the CMOS technology approaches its limitations, horizontal scaling in the form of parallel processing draws more attention to research delivering highest compute performance levels. In a few years we may observe the emergence of many-core on chip systems with as many as 256 cores and more. Consequently, as the number of cores increase, the interconnection of such many-core processors play the dominant role in imposing the performance boundaries of such systems. In addition, with the rise of computational-intensive applications, even in our portable electronics such as cell-phones, System-On-Chip (SOC) devices require to satisfy very aggressive latency, power, and area requirements. NoC paradigm is a promising alternative to the traditional communication solutions such as buses or direct point to point connection. In general, NoCs offer scalable modular interconnects along with low-latency and high bandwidth for the future 100+ core era. However, those mentioned advantages come at price of consuming up to more than a quarter of a chip power budget. That being said, it is desired to utilize the allocated resources to the NoC to their maximum potential.\\

Historically, NoCs are  designed to serve the cores they connect in a static fashion. Such static architectures result in full insensitivity to the changes of behaviors and patterns of communications within their environments, which itself leads to underutilization of a NoC. In other words, the mismatch between the latency and bandwidth requirements of applications and those offered by the static NoC translates into higher consumed power and longer delivery latencies.\\ 

A class of solutions to this problem offer to design NoCs that are tailored for one or more specific application(s) \cite{jalabert2004pipescompiler}\cite{bertozzi2005noc}\cite{murali2006designing}\cite{zhong2011application}. Another class of solutions propose addition of express links between the node with high hop counts\cite{grot2009express}\cite{kim2007flattened}\cite{narayana2017hyppi}. Express links may mitigate the underutilization problem and add bandwidth by reducing the average hop. But, NoCs may still be severely underutilized since the communication behaviors of applications may vary or may have operational phases due to factors such as changes in memory access patterns. Such changes of behavior explain the current trend shift in NoCs from static to dynamically reconfigurable designs.\\

In this paper we are proposing an adaptive dynamic reconfigurable NoC design. Our NoC adapts to changes in environment by taking measurements of environment and react to the measurements by augmenting the topology by an adaptable optical express bus. In fact, we also enable the dynamic adaptation of our measurement system in response to behavior changes of environment. To authors knowledge the latter approach had not been investigated prior to this study in the context of NoCs. It should be noted that the design space of reconfiguration is vast and beyond the scope of this study. The primary objective of this paper is to show the potential of adaptive dynamic measurement in addition to conventional reconfiguration techniques. Our design is motivated by the Dynamic Data Driven Application System (DDDAS) paradigm\cite{darema2004dynamic}. In DDDAS, computations and measurements form a dynamic closed-loop feedback in which they tune one another in response to changes in the environment. We expect the practice of applying DDDAS concept in NoC design would improve both performance and power efficiency.\\

In order to investigate the effectiveness of adaptive measurements in NoC performance, we simplify our design by enabling only a single any-to-any custom connection during run-time. Although straightforward in theory, this modification is not entirely cheap in practice and results is increased area footprint and consumed power. But, the simplification ensures that any performance improvements we achieve is mostly coming from the dynamic adaptive measurement rather than other NoC performance enhancement techniques. Having said that, we augment the topology by enabling only one optical express bus connection during different time window steps of our NoC operation. We also propose a reconfiguration algorithm based on gradient descent optimization technique to adjust the measurements at run-time, which we will explain in section \ref{sec:reconf} in details.

We describe the major contributions of this paper as follows,
\begin{itemize}
\item We propose an adaptive dynamic NoC design, which not only dynamically reconfigures the topology at run-time, but also dynamically tunes the underlying measurement process. Hence we form a closed-loop feedback of computations and measurements.
\item We investigate the effectiveness of our proposed dynamic measurement by simulating our design on real application kernels and synthetically generated traffic.
\item We compare our simulation results with a non-adaptive dynamic reconfigurable NoC in terms of latency and power.\\

The rest of this paper is organized as follows. In section \ref{sec:hyppi} we introduce and discuss the Configurable Hybrid Plasmonic Photonic Interconnect (CHyPPI). Section \ref{sec:d3} then goes over the overall architecture of D\textsuperscript{3}NOC and its parameters. In section \ref{sec:reconf} we introduce our adaptive measurement process and the reconfiguration algorithm. Evaluation results in terms of latency and power are given in section \ref{sec:results} and the behavior of measurement window in time is analyzed. Finally, in section \ref{sec:conclude} give our concluding remarks.
\end{itemize}

\section{Hybrid Plasmonic Photonic Interconnect (HyPPI)}
\subsection{Solving the Electronic Interconnect Bottleneck}
The electronic interconnect has been a critical bottleneck for on-chip communications and networks \cite{borkar3state}\cite{ujaldonlook}\cite{miller2009device}. As the needs of communication bandwidth and the number of parallel cores keep increasing, the latency and energy performance of electronic interconnects, which mainly determined by the resistance and the capacitance of the wire, is facing more and more challenges in long distance communication. Moreover, the heat dissipation in recent dense integrated electronic chips has become another issue that chip designers need to address.\\

Based on these, photonic interconnects are considered as a viable on-chip communication method since the mid-2000s due to its higher bandwidth as well as the wavelength division multiplexing (WDM) ability inspired by the intrinsic parallelism of bosons. However, due to the diffraction limit of light and the high quality (Q) factor based devices, such as microring resonators and ring modulators, the operating frequency and the energy efficiency are quite limited by the long photon lifetime within such devices. In addition, the wavelength picking process of these ring-based devices requires extra heating energy, which has been estimated to cause about 40\% of the entire static power of a link \cite{batten2009building}\cite{sun2012dsent}.\\

To solve this aforementioned problem, plasmonic devices are proposed to overcome the diffraction limited optical modes and achieve high field density. This addresses the footprint as well as the operating frequency and the energy efficiency problems by providing enhanced light matter interaction (LMI)\cite{fratalocchi2015nano}\cite{pickus2013silicon}\cite{sun2015case}\cite{ma2015indium}\cite{li2015monolithic}\cite{narayana2017morphonoc}\cite{sun2016universal}. However, due to the intrinsic Ohmic loss, using plasmonic waveguide for signal propagation is extremely lossy. Thus, a hybrid interconnect solution has been proposed recently termed HyPPI that combines passive photonic waveguide only for signal propagation and active plasmonic devices to provide fast and energy efficient light manipulation \cite{treyz1991silicon}.

\subsection{Modulation Strategy}

Depending on the modulation strategy, HyPPI can be categorized into intrinsic modulation and extrinsic modulation. The main difference of these two is the modulation part of the link. For an extrinsic HyPPI, an external modulator is used to load the electrical data onto light signal. And for an intrinsic HyPPI, no modulator is needed and the signal will be generated directly from the laser or other light sources. After analyzing our mesh network topology, intrinsic HyPPI has been chosen to provide the best fit since modulator bypassing losses can be eliminated for long range communication between cores.\\

However, on-bus modulator introduces substantial insertion losses even while inactive. Therefore, an off-bus modulator is necessary for HyPPI before it can be deployed on NoCs. Additionally, off-bus light detection is also important. Although a photodetector can be directly placed at the end of a waveguide bus for a point-to-point link, it would be quite inefficient in a network structure with multiple nodes/hops.  Alternatively, the waveguide needs to be segmented with O-E-O conversions at every point on the waveguide where we need configurability and thus a photodetector/modulator pair.  

\begin{figure*}[ht]
\centering
\subfloat[Overview of mo-detector-based link]{
   \includegraphics[width=0.5\linewidth]{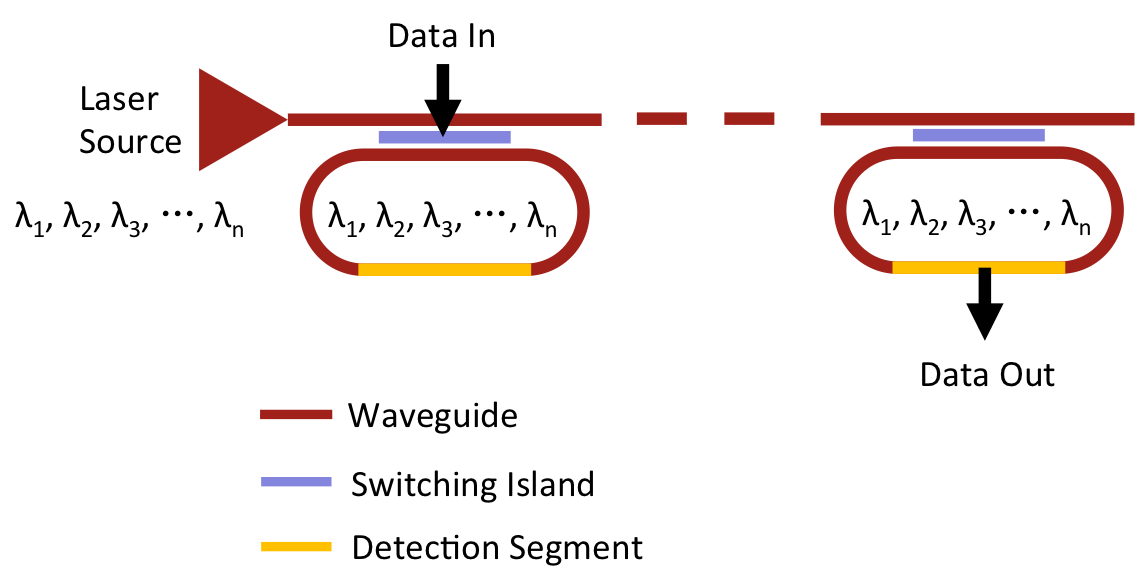}
   \label{subfig:modetoverview}
   }
\subfloat[Mo-detector Operation]{
   \includegraphics[width=0.45\linewidth]{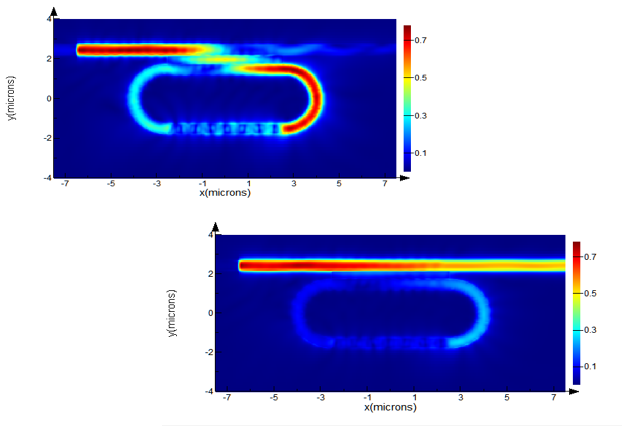}
   \label{subfig:modetoperation}
  } 
  \\
\subfloat[2$\times$2 Plasmonic Switch]{
   \includegraphics[width=0.45\linewidth]{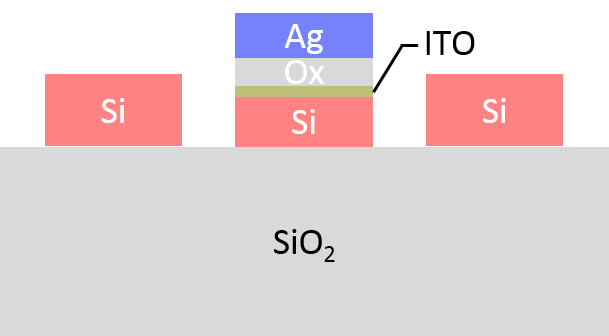}
   \label{subfig:2x2switch}
  }  
\subfloat[Photodetector Integration]{
   \includegraphics[width=0.45\linewidth]{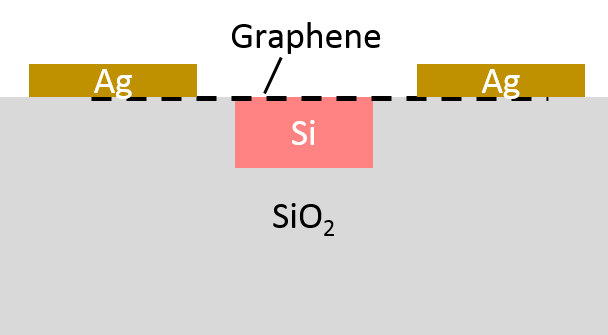}
   \label{subfig:photodet}
  }
\caption{Plasmonic Mo-detector Device for Configurable HyPPI}  
\label{fig:modetector}
\end{figure*}

\subsection{Mo-detector}
To achieve off-bus modulation and detection, we propose a new device structure call Mo-detector which combines the function of a modulator and a detector, see   Fig.~\ref{subfig:modetoverview}. The modulation function is integrated on the racetrack ring waveguide controlled by a 2$\times$2 plasmonic switch (Fig.~\ref{subfig:2x2switch}) based on HPP which was proposed by Ye et al. ~\cite{ye2015compact}. By applying a voltage on the metal contact, the refractive index of Indium Tin Oxide (ITO) would change from nOFF = 1.042-0.723i to nON = 1.964-0.001i based on the experimental results from the literature~\cite{oulton2009plasmon}.\\

By changing the bias voltage of the ITO layer, its effective index changes, which will also affect the coupling length at different states. Since the switch length just equals to the coupling length of the \emph{cross} states with no voltage applied (light couples from the bus to the ring), the light from the bus will be directly coupled to the switch island and then into the ring (Fig.~\ref{subfig:modetoperation}, top part). After applying the bias voltage, the coupling length of the \emph{bar} state (light reflects back to the bus again) will be much longer, and there will be not enough space for the light to couple. Therefore, the light will be bounced back and stay within the bus (Fig.~\ref{subfig:modetoperation}, lower part). We chose to use this hybrid plasmon polaritons (HPP) plasmonic switch for two reasons. First, its compact structure greatly enhanced the light-matter-interaction (LMI) and therefore the device length is energy and area efficient. Secondly, it is off the bus waveguide and has good electrical controllability.\\

The other function of this device is light detection. In order to achieve this function, any compatible light absorption photodetector that can be directly integrated onto a silicon waveguide could be applied to this Mo-detector (Fig.~\ref{subfig:photodet}). Here, we borrowed the ultrafast graphene photodetector proposed by Gan et al.~\cite{gan2013chip}. Based on the results they reported, this metal-doped graphene junction coupled to the waveguide has more than 0.1 A/W responsivity with high operating speed and wide bandwidth from 1450 to 1590 nm. Note, this responsivity performance is achieved under zero-bias voltage. Therefore, this device is able to extract data and convert into electrical signals at detection state.  Additionally, it also absorbs the light coupled into the ring in order to generate a "0" signal (binary signal) during modulation.\\

The overall performance of this Mo-detector device makes it a good choice for NoCs. The insertion loss is as low as 0.08 dB, which means the transmission loss is negligible when bypassing this device. On the other hand, the On-Off-Ratio (the power ratio between modulating a "1" and a "0", also called \emph{extinction ratio}) of this device is as high as 15.53 dB, which provides quite clear signals for detection. We refer to this new, Configurable HyPPI as CHyPPI.  

We list some critical parameters of the optical parts we used in the network simulation in table \ref{tab:parameters}.
\label{sec:hyppi}
\begin{table}[t]
\centering
\begin{tabular}{cl|c}
\hline
\hline
\multicolumn{2}{c|}{Parameters} & CHyPPI \\
\hline 
\hline 
\multirow{2}{*}{Laser} & Efficiency (\%) & 20 \\
 & Area ($\mu$m\textsuperscript{2}) &  0.003 \\
\hline 
\multirow{7}{*}{Modulator} & Speed (Gbps) &  3470 (50) \\
   & Energy Efficiency (fJ/bit) &  2.59 \\
   & Insertion Loss (dB) & 0.08 \\
   & Extinction Ratio (dB) &  15.53 \\
   & Area ($\mu$m\textsuperscript{2}) & 32 \\
   & Capacitance (fF) &  0.58 \\
   & Bias Voltaage (V) &  2$\thicksim$3 \\
\hline
\multirow{4}{*}{Photodetector} & Speed (Gb/s) & 50/700 \\
& Energy Efficiency (fJ/bit) & 0 \\
& Responsivity (A/W) &  0.1 \\
& Area ($\mu$m\textsuperscript{2}) & 4 \\
\hline
\multirow{4}{*}{Waveguide} & Propagation Loss (dB/cm) &  1 \\
  & Pitch ($\mu$m) &  5 \\
  & Width ($\mu$m) &  0.45 \\
\hline 
\end{tabular}
\caption{ CHyPPI link parameters}
\label{tab:parameters}
\end{table}

\section{D\textsuperscript{3}NOC Design} 
\begin{figure}[htb]
\centering
   \includegraphics[width=\linewidth]{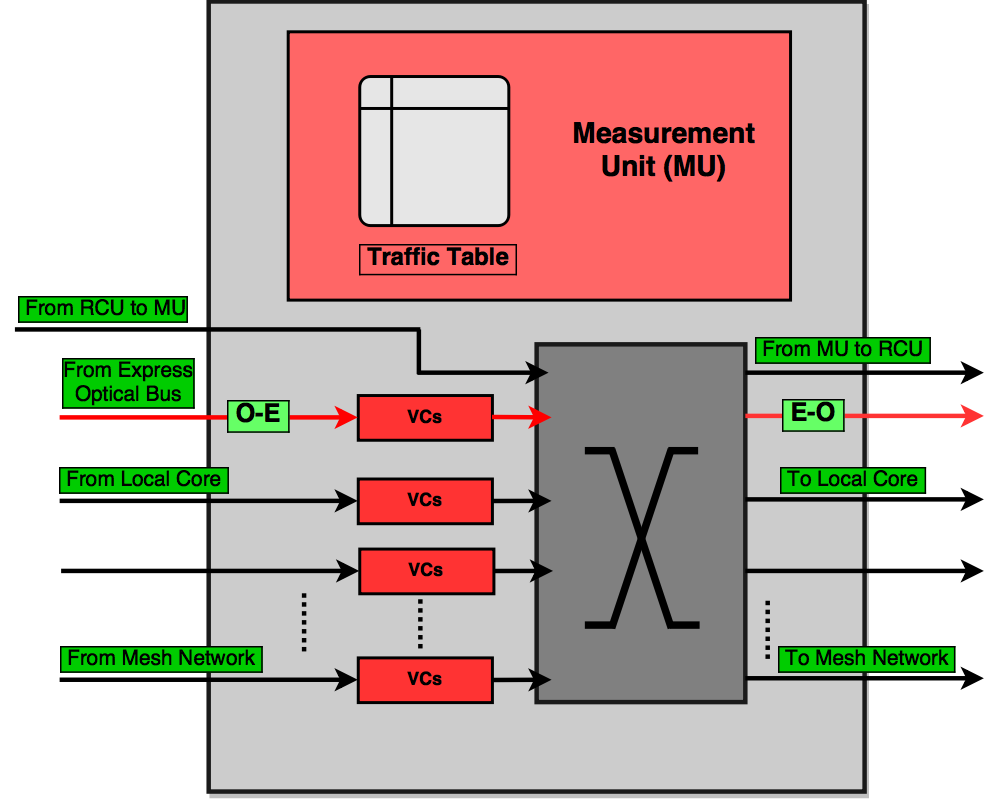}
\caption{Hybrid router architecture}  
\label{fig:router_arch}
\end{figure}

Our design is built on a $16\times 16$ base 2D-mesh structure with cores spaced $1mm$ apart. As mentioned earlier, he optical express bus is formed by connecting all the nodes through a serpentine connection as depicted in Figure 
\begin{figure}[htb]
\centering
   \includegraphics[scale=0.045]{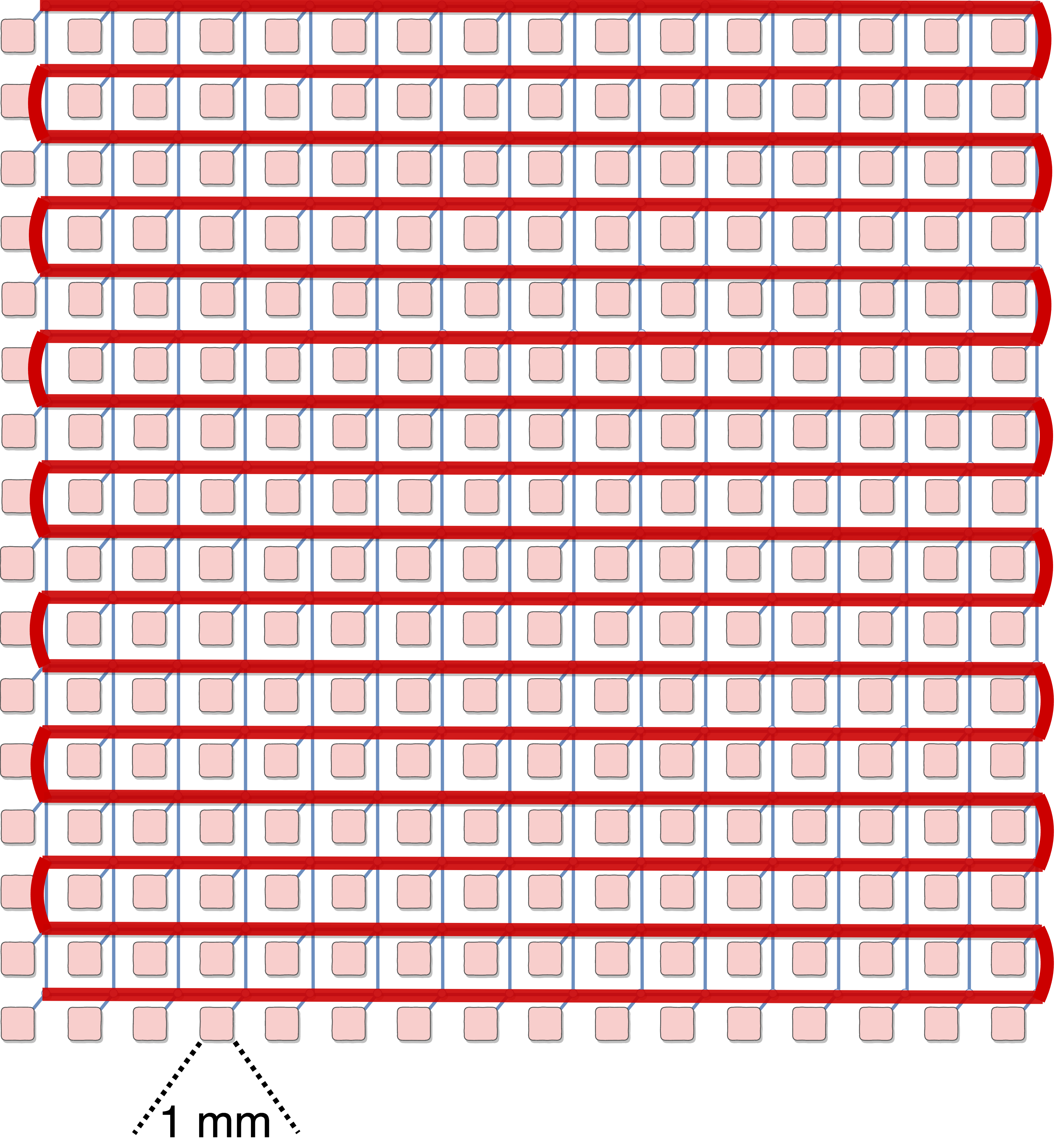}
\caption{A serpentine-like structure connects all the cores in a 16$\times$16 NoC. Each node is equipped with a mo-detector and its laser source.}  
\label{fig:d3noc}
\end{figure}
The NoC runs a at a clock frequency of 0.78 GHz. This reason behind picking this particular clock frequency is due to the fact that HyPPI links only support a single wavelength. In addition, the the electronic drivers of the link operate at 50Gb/s. As a result, a 50Gb/s HyPPI link and a flit size of 64 bits determines a 0.78125 GHz. It should be noted that we decided to have all our links to run at the same rate to avoid excess buffering due to frequency mismatches. We used DSENT \cite{sun2012dsent} to simulate power and area estimates for $11nm$ technology. \ref{tab:noc_parameters} shows the NoC parameters we used in this paper. Based on our simulation results, link traversals through electronic links incur a 1-clock-cycle latency. This number for an optical link is 2 clock cycle considering the optical to electrical conversion time at the receiver. The electrical to optical time delay at the transmitting router is included in the staging step of the buffer of transmitting router. We used conventional 5-port electronic routers for the base electrical mesh NoC and 7-port routers for D\textsuperscript{3}NOC as shown in Figure \ref{fig:router_arch}. The router connects to the express optical bus through one of the router ports. The last router port is reserved for reconfiguration communications as will be explained is section \ref{sec:reconf}.

\begin{table}[!h]
\centering
\begin{tabularx}{\columnwidth}{cl|p{3cm}}

\hline
\hline
\multicolumn{2}{c|}{Parameters} & Value \\
\hline 
\hline 
\multirow{3}{*}{Nodes} & \# Nodes & 16$\times$16 (256 nodes) \\
                       & Core Spacing & 1 mm \\
                       & Core Clk Freq   & 0.78125 GHz \\
\hline                        
\multirow{8}{*}{Router} & Flit Size & 64 bits \\                    
                        & \# Ports & 5 (base) or 7 (hybrid) \\
                        & \# Virtual Channels (VCs) & 4 \\
                        & Buffers per VC & 8 flits \\
                        & Pipeline Depth & 3 stages \\
                        & Area & From Modified-DSENT \\
                        & Static Power & From Modified-DSENT \\
                        & Dynamic Energy per Flit & From Modified-DSENT \\ 
\hline 
\multirow{5}{*}{Link}  & Latency & 1 clk for Electronic, 2 clks for HyPPI\\ 
                       & Capacity & 50 Gb/s \\ 
                       & Area & From Modified-DSENT \\ 
                       & Static Power & From Modified-DSENT \\
                       & Dynamic Energy per Flit & From Modified-DSENT \\ 
\hline 
\end{tabularx}
\caption{ Network Parameters Used for NOCs in This Work}
\label{tab:noc_parameters}
\end{table}
\label{sec:d3}

\section{Reconfiguration}
To achieve adaptive reconfiguration, we bridge over the cores with high traffic volume that are located farther away from one another by lending them the optical express bus. We define two important time interval namely, reconfiguration period $(\Delta T_{rp})$ and operation window $(\Delta T_{ow})$. We may use the term "measurement window" and "operation window" interchangeably in the rest of this paper because they both refer to the same time interval. Reconfiguration period is a constant time period during which the NoC reconfigures from one topology to another. During the reconfiguration period future owners (source-destination cores) of the bus are identified and notified. The bus is released from previous owners and is allocated to the next owners. On the other hand, during the operation window, NoC continues it normal routine operations while cores keep track of the amount of traffic they communicate with other cores.\\

A global Reconfiguration Control Unit (RCU) acts as reconfiguration controller. Towards the end of each operation window nodes send a message to the RCU notifying the RCU of the their own ID, the ID of the core they communicated the most with, the volume of communication with that particular core, and the total amount of communication they had in number of flits during the previous operation window. Then, RCU enters the reconfiguration period during which it: (a) determines those cores that express optical bus will be allocated to for the next operation window; (b) calculates the length of next measurement window, and (c) broadcasts the IDs of the source and destination cores that will hold the express optical bus during the next time window to all the cores. Please note that all the core require to know which two cores will take hold of the express optical bus during the next time window in order to apply control signals for acquiring/releasing the bus and also to update their routers' routing paths. \\

In our proposed NoC the length of the measurement window adapts to environment so that to minimize an specified cost function. Currently, we have defined the cost function as the average latency over the whole nodes of the network. Having said that, it may be preferred to assign a different cost function based on other priorities such as link utilization, power consumption, etc.\\

We adopted gradient descent to adjust the length of measurement window. In general, gradient descent algorithms go thorough three phases namely: (a) determine the descent direction (whether the window needs to be longer or shorter), (b) select a step size, and (c) update the previous weight (window size). In our case the gradient value, which is essentially a single first order difference, is found by the following by,
\begin{equation}
Grad_{\Delta (t)}=\frac{Latency(\Delta T_{ow(t)})-Latency(\Delta T_{ow(t-1)})}{\Delta T_{ow(t)}-\Delta T_{ow(t-1)}}
\end{equation}
where $Latency(\Delta T_{ow(t)})$ is the total network latency during the current Operation Window, $Latency(\Delta T_{ow(t-1)})$ is the total network latency during the previous time window, and $\Delta T_{ow(t)}$, $\Delta T_{ow(t-1)}$ are the length of current and previous time window respectively. Then, we update the Operation Window using the following step size,
\begin{equation}
\Delta T_{ow(t+1)} = \Delta T_{ow(t)}- (\alpha\times Grad_{\Delta (t)})
\end{equation}
where $\alpha$ is a fraction value between 0 and 1 that allows the manual control of step size on top of the gradient. During our initial simulation we realized that we need to bound the resizing rate of the measurement window. Our initial measurements showed that for measurement window sizes of less than 100 cycles, the cost of reconfiguration would become significant and thus resulting in higher latencies. The latter constraints in mainly imposed by the cost of reconfiguration window. During the reconfiguration window the NoC is only dedicated to moving the reconfiguration traffic around. As a result, the quicker the measurement window changes, the more the length of reconfiguration window becomes comparable to length of measurement window. In addition, small window sizes would become over-sensitive to small changes in the traffic patterns so that we would again pay the reconfiguration cost for transient patters. These two ultimately defined the lower bounds of our measurement window size. On the other hand, if we allow for sudden unbound expansion of measurement window size, windows may suddenly expand orders of magnitude. Such unbound expansions would lead to total insensitivity to traffic patterns. In our simulations we observed that in almost all cases, after a sudden expansion of measurement window size, the NoC would be forced to operate for a long time without any reconfiguration. In fact, for most cases the NoC never gets to recover from a long measurement window because the window would last towards the end of simulation. As a result, we restricted the expansion of measurement window to 10 times of previous window size at each step.

\begin{algorithm}
    \caption{Reconfiguration Algorithm}
  \begin{algorithmic}[1]
    \REQUIRE Reconfiguration Period $\Delta_{RP}$ to start
    \INPUT List of the volume of total communication, and the core they most communicated with, and amount of that particular communication for each core in the network
    \OUTPUT The two cores that will own the express optical bus for the next $\Delta T_{OP}$ and the size of next $\Delta T_{OP}$
    \STATE Find the two cores with the max communication
    \STATE Calculate the overall latency for current $\Delta T_{OP}$
    \STATE Calculate the next $\Delta T_{OP}$ length.
    \STATE Broadcast the next $\Delta T_{OP}$ length to all the cores.
    \STATE Send control signal to the 2 cores owning the express optical bus for the future $\Delta T_{OP}$
  \end{algorithmic}
\end{algorithm}
\label{sec:reconf}
\section{Evaluation and Results}
In this section we first present our simulation setup and then the simulation results. The goal of this study is to demonstrate potentials of adaptive dynamic measurement systems coupled with dynamic topology reconfiguration in the in NoCs. Thereby, we compare the performance and cost of our design with: (a) a base fully electrical NoC, and (b) with a topology-reconfigurable NoC the same as our design but with a static measurement system.

\subsection{Evaluation Setup}
Historically, NoC simulators not only do not allow for reconfiguration of a measurement unit of NoCs at run-time, but also they are capable of changing the topology at run-time as well. Therefore we developed a custom-built latency and power simulators.\\

This simulator accounts for a single cycle link traversal time for electronic links. The electronic routers impose two clock cycle delays for packets through them. For each source and destination travel we also apply 1 cycle cycles for local core to router and router to local core each. Per our calculation the total express optical bus  traversal time falls within a single clock cycle. In addition, each packet going through express optical bus incurs 1 more clock cycle delay due to O-E conversions at destination. The delay of E-O conversion at the source can be ignored since the conversion can be done in parallel within the same clock cycle of source router delay.\\

We applied a modified version of X-Y routing, which we call X-Y* routing. The X-Y* routing is similar to X-Y routing except that if a packet enters a core, which owns the express optical bus to packets destination core, the packet would take the bus to destination. While conventional X-Y routing would reserve the express optical bus for the use of packet with the same source and destination of express optical bus, X-Y* routing allows more number of traveling packets in the network to take advantage of the an established express optical bus. The X-Y* routing is deadlock free since we are not introducing any circular paths. Please note that we could have used a more expensive routing algorithm such as shortest-path routing. However, because we modify the topology during each operation window, routers would require to calculate the shortest path for each operation window. The latter not only imposes a more costly router, but also the calculation of shortest path for each operation window if feasible, would be very expensive.\\

We generously set the length of our reconfiguration to be 50 clock cycles. Per our calculations 50 clock cycles allow for cores from edges of the NoC to report their traffic data over the previous operation window to the RCU at the center of NoC and RCU to find the next owners of the express optical bus and find the next operation window size. It should be noted that the size of the traffic report from the core to RCU fits an 8byte packet (4 bytes for overall flit count of the core, 4 bytes for the flit counts to the most communicated node, and 2 bytes for the address of that particular core). Thus, We expect the process of cores reporting to the RCU should be easily fit within reconfiguration period and leave enough time for RCU calculations, which is implemented in hardware.\\ 

The performance of our 256-core NoC was evaluated against Class A kernels of the NAS parallel benchmarks (NPB) suite\cite{vanderwijngaart2003parallel}, namely EP, CG, FT, LU, and MG. Traces were obtained by executing the kernels on an in-house Cray XE6m supercomputer using MPICL. We also evaluated our design against the synthetic traces generated by the PacketGenie tool \cite{packetg}. The two category of traces we used include frequently communicating pair (fcp) and many to few to many (mfm).

\subsection{Simulation Results}
\begin{figure}[!h]
\centering
   \includegraphics[width=\linewidth]{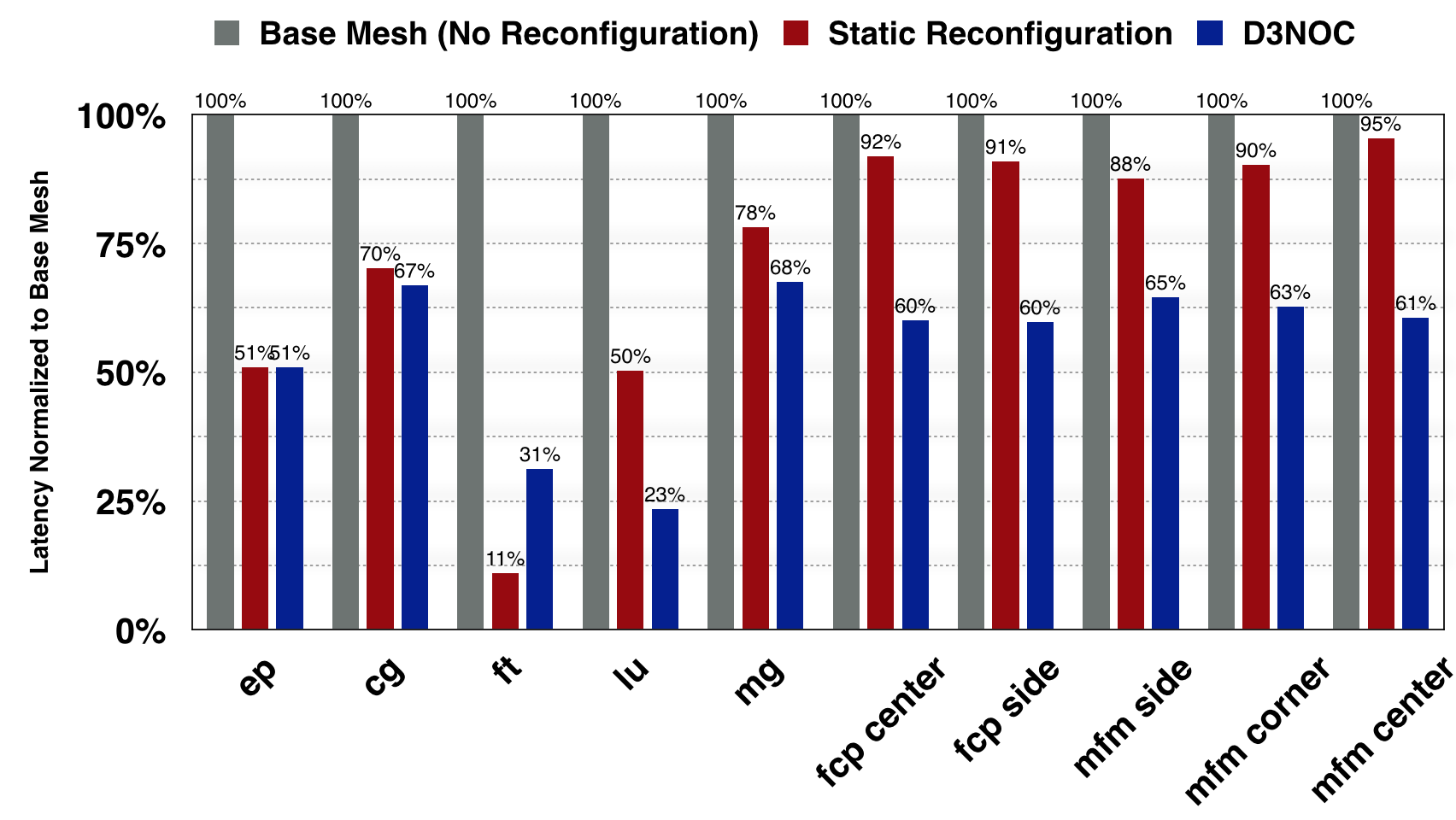}
\caption{Latency simulation results for NAS Parallel Benchmarks and synthetic traffic generated using PacketGenie suite.}  
\label{fig:Latency}
\end{figure}

\begin{figure}[htb]
\centering
   \includegraphics[width=\linewidth]{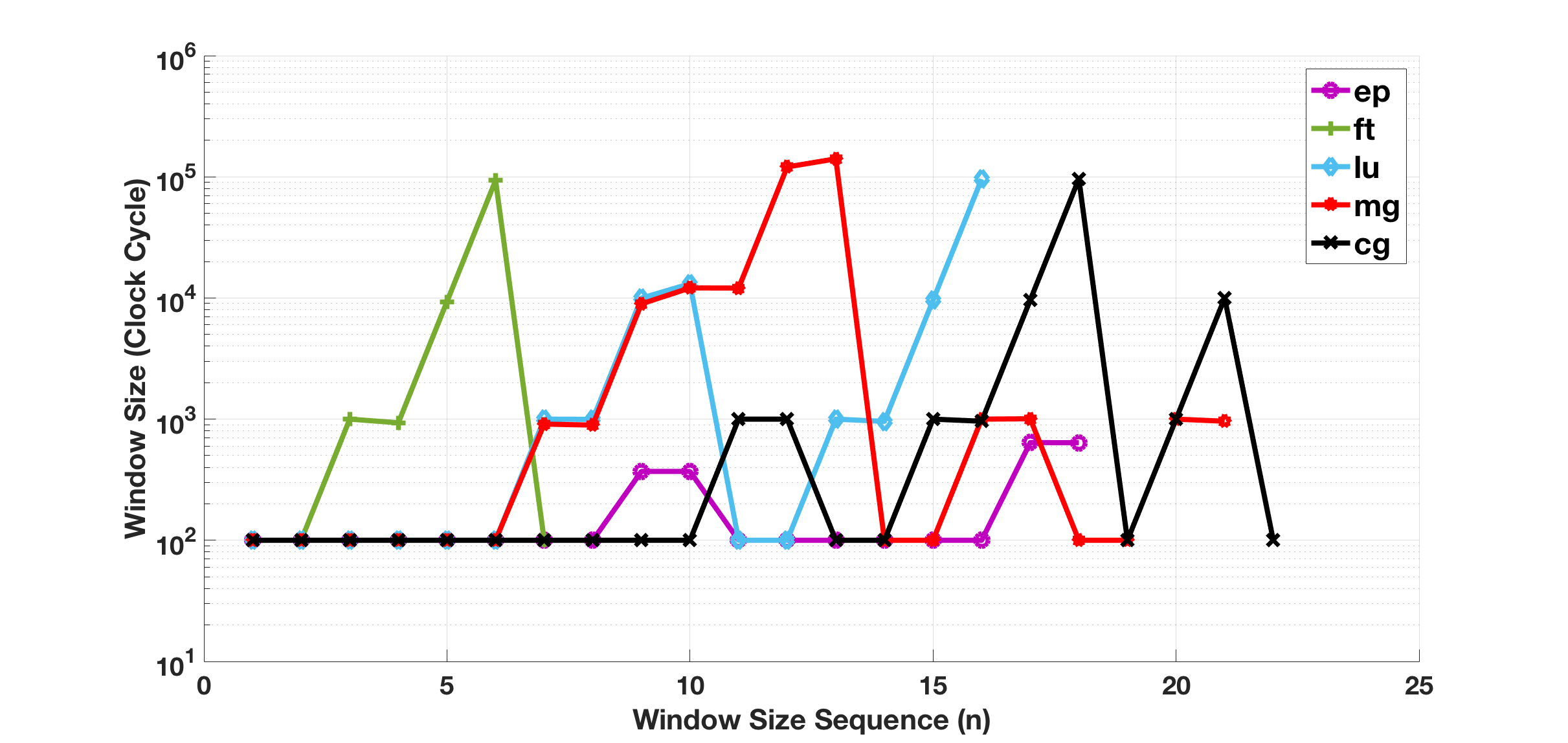}
\caption{Progression of measurement window size over time for various kernels of the NAS parallel benchmark. The horizontal axis is representative of sequence of change. The vertical axis represents the measurement window size in Cycles.}  
\label{fig:grad_nas}
\end{figure}

Figure \ref{fig:Latency} depicts the summarized results of the latency for both the NAS application suite and synthetic traffic generated by PacketGenie tool. We normalized the recorded latencies for each traffic pattern and for each NoC architecture to that of the base electrical mesh. Our results show that the latencies for all synthetic traffic pattern improved by 5\% to 12\%  for a static reconfiguration with a fixed measurement window. However, the same numbers for the adaptive window show a more significant improvement up to 40\%. On the other hand the latency improvements for the NAS parallel benchmarks show a different behavior. FT kernel which has all-to-all type of traffic benefits the most from reconfiguration. For FT kernel the static measurement window outperforms the adaptive measurement window. CG comprises of short range traffic and thus benefited relatively less from reconfiguration compared to other kernels except MG. For MG we expected to observe the most latency improvements since the major portion of MG consist of long range traffic. In contrast, LU, which is significantly comprised of 1-hop traffic shows latency improvements of 50\% for a static measurement window and 77\% with an adaptive measurement window.  \\
\begin{figure}[!h]
\centering
   \includegraphics[width=\linewidth]{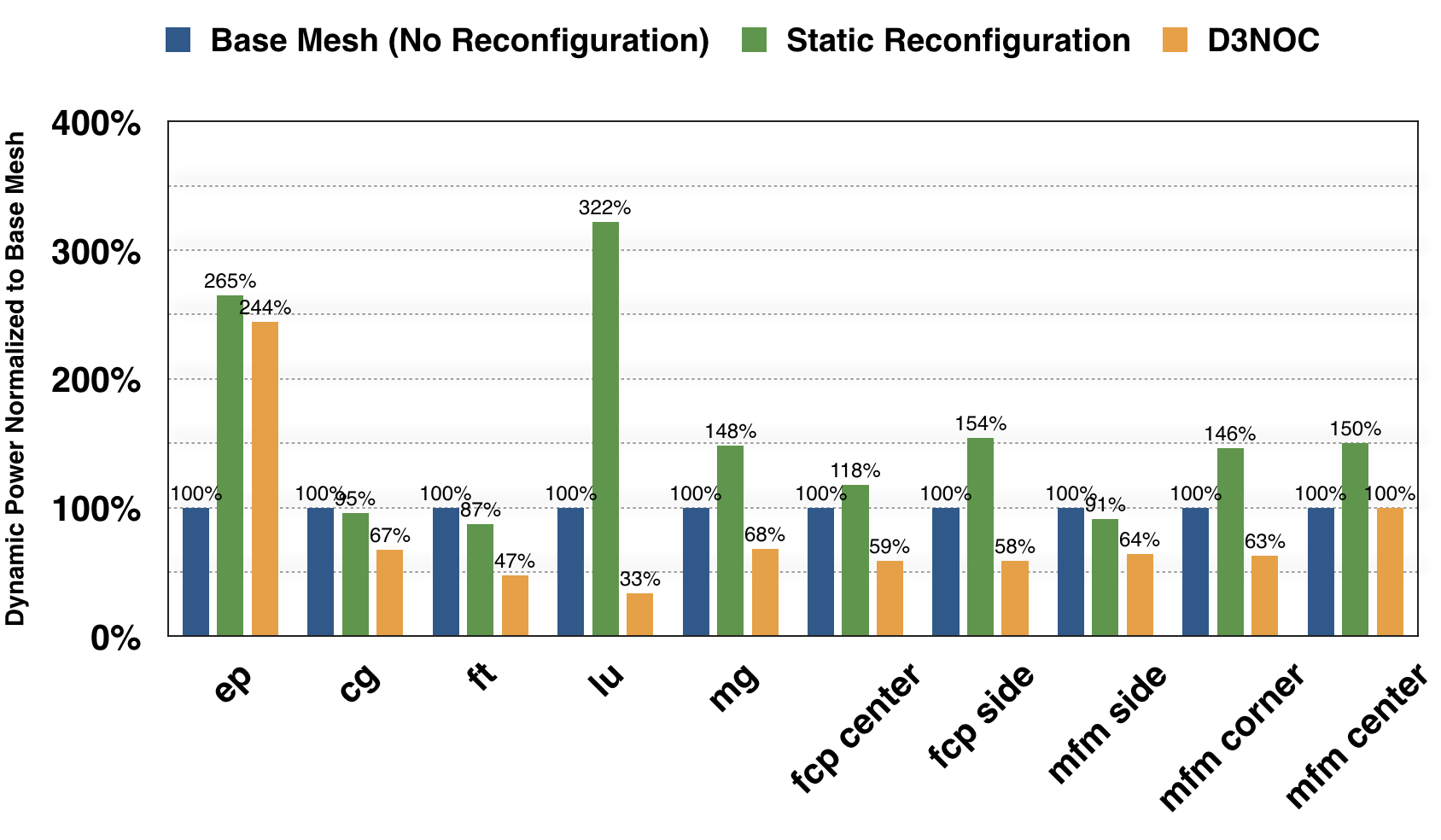}
\caption{Dyanmic power simulation results for NAS Parallel Benchmarks and synthetic traffic generated using PacketGenie suite.}  
\label{fig:power}
\end{figure}
On the hand, dynamic power simulations showed a different behavior. In our design with CHyPPI link, mo-detectors, and intrinsic laser sources, the light travels only the distance it requires (between source and destination in all cores). Therefore, energy consumed per flit in each window will be different. The reconfiguration with static window size resulted in increased power across the board except for FT and CG kernels. Both FT and CG have mid range, but steady communication traffic patterns. Long-haul communications through express optical bus may result in excess power consumption, more specifically if the bus is underutilized due to non-adaptivity of measurement window. The results show that except for EP and mfm center, adaptive measurement window lead to power saving improvements. EP kernel is a very short kernel during which cores initialize and send a few packets around. As a result EP may not be an ideal benchmark for characterization. We also used DSENT tools to estimate the area overhead of D\textsuperscript{3}NOC. Area estimation showed 45\% increase for dynamic NoC with static window and 52\% for D\textsuperscript{3}NOC Design due to measurement overhead. It should be noted that we used active area as a measure of comparison. Besides latency and dynamic power, we also examined the changes of measurement window size. Figure \ref{fig:grad_nas} shows how measurement windows change over time.  

We were able to associate some of the behaviors of the measurement window size with the prior knowledge of the kernels. For instance, the MG and LU kernels follow the same pattern in the beginning, but the measurement window size of the LU rolls off and while that for MG continues to increase. The latter may correspond to the fact that, LU and MG kernels show a rapid growth of aggregate communications in the beginning of their execution. After reaching that peak aggregate communication, both kernels' communication volume start to descend. Later, MG goes through multiple cycles of similar communications, but LU does not. On a different note, FT quickly picks up to a series of regular high-volume aggregate communication and follows the same pattern throughout its execution. We believe that may be the reason behind the early expansion of measurement window.\\

\begin{figure}[htb]
\centering
   \includegraphics[width=\linewidth]{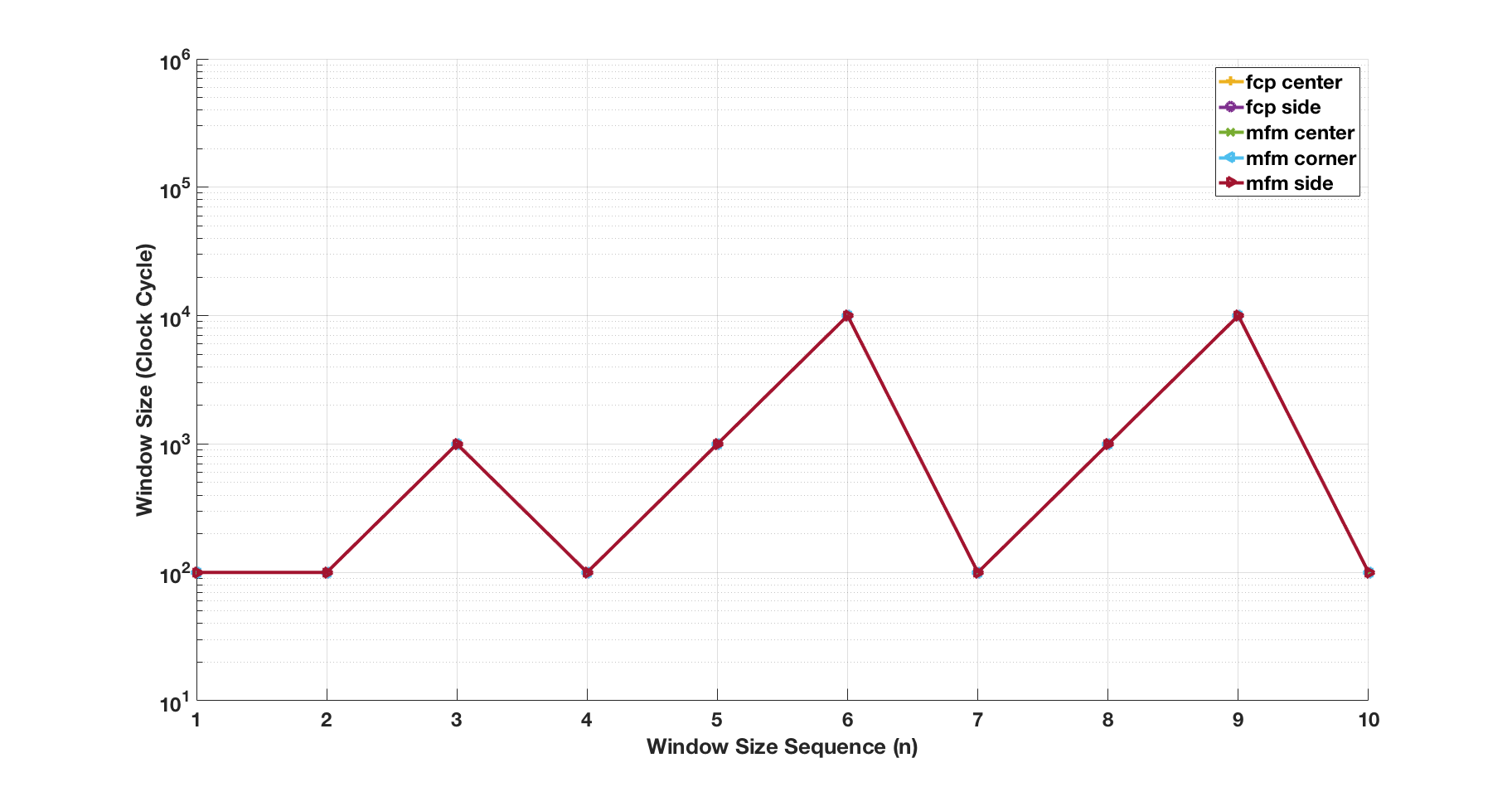}
\caption{Progression of measurement window size over time  for synthetically generated traffic patterns. The plot shows the exact same progression trend of the measurement window size.}  
\label{fig:grad_syn}
\end{figure}

We generated the same plot of measurement window size progression for our synthetic traffic as shown in Figure \ref{fig:grad_syn}. Notably, all the curves exactly align on top of each other. 

To explain the identical window progression trend, we believe the underlying random process that generated the trends is the same for all synthetically generated traffic. All generated traffic traces are 400k cycles long. It also seems that no matter which set of nodes are communicating, the aggregate traffic behaviors are the greatly similar for these traces, thus resulting in corresponding same patterns in adaptation of measurement windows.
\label{sec:results}

\section{Conclusion}

In this paper we conceptualize and investigate a novel reconfigurable NoC termed the D\textsuperscript{3}NOC. The design is motivated by Dynamic Data Driven Application System (DDDAS) paradigm. The goal of this study is to show the potentials of adaptive dynamic measurement in conjunction with current common reconfiguration techniques, such as reconfiguration of topology. In D\textsuperscript{3}NOC not only we reconfigure the topology of our NoC at run-time, but we also adaptively adjust the measurement process leading to the original reconfiguration. In addition, we took advantage of optical interconnects as an express bus to augment the topology of our NoC. Since conventional electronics was incapable of providing us with such a long, low-latency bus, and nano-photonic links were too power inefficient in addition to their large footprint, we decided to use configurable hybrid photonic-plasmonic links. We evaluated our design against kernels of the NAS Parallel Benchmarks in addition to synthetically generated traffic using PacketGenie tool. The simulation results showed dynamic adaptation of measurement process can lead to further improvements in latency and dynamic power on top of improvements gained over dynamic reconfiguration but with static measurements.
\label{sec:conclude}

\addtolength{\textheight}{-10.8cm}   




\newcommand{\BIBdecl}{\setlength{\itemsep}{0.15 em}}

\bibliographystyle{IEEEtran}
\bibliography{nocBIB}

\end{document}